\documentclass{mn2e}
\title [Bias in the use of parallaxes] {Bias in Absolute Magnitude
Determination from Parallaxes}
\author [Michael Feast] {Michael Feast\\
Astronomy Department, University of Cape Town, Rondebosch, 7701, South Africa.\\mwf@artemisia.ast.uct.ac.za}
\begin{document}
\maketitle
\begin{abstract}
Relations are given for the correction of bias when mean absolute magnitudes
are derived by the method of reduced parallaxes. The bias in the case of the
derivation of the absolute magnitudes of individual objects is also
considered.
\end{abstract}
\section{Introduction}
The general methods for dealing with bias in the analysis of astronomical
(or other) data were set out by Eddington (1913, 1940, see also Dyson 1926).
Athough these are quite straightforward, the current literature shows that
there is some uncertainty and misunderstanding in these matters. the present
paper attempts to clarify the situation in the case of the derivation of
absolute magnitudes from parallaxes.

Since the Hipparcos catalogue (Perryman et al. 1997) became available there
has been increased use of parallaxes for absolute magnitude determination and
future missions such as GAIA will lead to further work in this area. It is
therefore important that there should be agreement on the question of
bias correction.

If objects are selected for analysis on the basis of their observed
parallaxes ($\pi$) or weighted (including selection or rejection) by the
ratio of $\pi$ to its standard error, $\sigma_{\pi}$, then the result is 
subject to bias (Lutz \& Kelker 1973). On the other hand, provided there is no
selection or weighting according to $\pi$ or $\pi/\sigma_{\pi}$ and if the 
relative absolute magnitudes of the objects in a group are known then the 
method of reduced parallaxes can be used to convert relative values to
absolute. In the next section the biases that can occur in this method are
set out. Following this, the question of the determination of the absolute
magnitudes of individual objects from their parallaxes will be considered.

\section{The method of reduced parallaxes}
There have been a number of discussions of bias applicable to the method of 
reduced parallaxes but none dealing entirely satisfactorily with the
situation likely to arise in practice. For instance Turon \& Cr\'{e}z\'{e}
(1977) consider only the case when all the objects have identical absolute
magnitudes (i.e. there is no dispersion in absolute magnitude) and the
discusion of Ljunggren \& Oja is not completely relevant to the present
purpose and seems to confuse the intrinsic dispersion in absolute magnitude
with the uncertainty in the mean absolute magnitude derived from parallaxes.
It should be noted that they, and the analysis of Malmquist (1920) to which
they refer, consider only the case when all the objects of the relevant
class down to a certain apparent magnitude are measured for parallax in the
area of sky considered.

In the case of recent discussions of the Hipparcos parallaxes of Cepheids
(e.g. Feast \& Catchpole 1997 (=FC), Oudmaijer et al. 1998, Groenewegen
\& Oudmaijer 2000) various views on bias corrections have been put forward,
none entirely satisfactory. In that particular case it is now generally agreed 
that
any correction is negligibly small (FC, Groenwegen \& Oudmaijer 2000,
Lanoix et al. 1999) and this is confirmed by Monte Carlo simulations
(Pont 1999). However occasions may well arise when these corrections are
significant.

In the method of reduced parallaxes it is assumed that the objects in the
group under discussion have not been selected by their measured parallaxes
$(\pi)$ or the ratio, $\pi/\sigma_{\pi}$. Consider first the case of
objects with a mean absolute magnitude per unit volume, $M_{o}$, and an
intrinsic dispersion $\sigma_{M_{o}}$. The method of reduced parallaxes
may be written:
\begin{equation}
\overline{10^{0.2M}} = \sum {0.01 \pi 10^{0.2 m_{o}}. p} / \sum {p}
\end{equation}
where the parallax is measured in milliarcsec, $m_{o}$ is the absorption
free apparent magnitude and $p$ is the weight given by;
\begin{equation}
(0.01\sigma_{T}. 10^{0.2m_{o}})^{2} = 1/p
\end{equation}
and where $\sigma_{T}$ is given by;
\begin{equation}
\sigma_{T}^{2} = 
\sigma_{\pi}^{2} + b^{2} \pi_{M_{o}}^{2}(\sigma_{m_{o}}^{2} + 
\sigma_{M_{o}}^{2}).
\end{equation}
Here,\\
$b = 0.2log_{e}10 = 0.4605,$\\
$\pi_{M_{o}}$ is the photometric parallax of an object computed from
$(m_{o} - M_{o})$.\\
$\sigma_{m_{o}}$ is the standard error of the reddening corrected apparent
magnitude.\\
Equation 3 is from Koen \& Laney (1998).\\
Equation 1 gives an unbiased estimate of $10^{0.2M_{o}}$ only if
$\sigma_{M_{o}}$ is zero.

Put $x = (m_{o} - M_{o})$. This will differ from the true distance modulus
by $\epsilon$ (say), due to observational errors in $m_{o}$ and intrinsic
dispersion in $M_{o}$. It is then evident that equation 1 yields an estimate
of;
\begin{equation}
\overline{10^{0.2M}} = \overline{10^{0.2(M_{o} + \epsilon)}} =
\overline{e^{bM_{o}}}. \overline{e^{b\epsilon}}.
\end{equation}

Consider objects all of the same $m_{o}$ (and $x$). From Malmquist (1920)
or more compactly, from Feast (1972) where some of the present 
nomenclature and sign convention is used, we find;\\
\begin{equation}
\overline{e^{b\epsilon}} = e^{0.5b^{2}\sigma_{t}^{2}} v(x-b\sigma_{t}^{2})/v(x)
\end{equation}
where;
\begin{equation}
\sigma_{t}^{2} = \sigma_{m_{o}}^{2} + \sigma_{M_{o}}^{2}
\end{equation}
and $v(x)$ is the frequency distribution of $x$ which would have been found
if a complete survey had been made. Note that this is the case whether or not
the objects under consideration actually form a complete survey.

Evidently at any $m_{o}$ an unbiased estimate of $\overline{10^{0.2M_{o}}}$
is obtained by multiplying the r.h.s. of equation 1 by the reciprocal
of the r.h.s. of equation 5. Summed over all $m_{o}$, as in a practical
case, the final result will depend $v(x)$ which in
general will depend on $x$. Furthermore if one were analysing parallaxes
over a significant galactic volume (as may well be possible with GAIA data),
$v(x)$ might not be the same in all heliocentric directions. In such cases
the problem would benefit from Monte Carlo simulations (Pont 1999,
Sandage \& Saha 2002). However if the underlying density distribution
is constant then the r.h.s of equation 5 becomes
$10^{-2.5b^{2}\sigma_{t}^{2}}$ (see for instance Feast (1972) equation 9).
Thus provided $\sigma_{t}$ is constant for the objects studied, this is
simply a constant bias factor.

The relation between natural and logarithmic quantities shows that if
$\sigma_{1}$ is the standard error of the mean value of $M_{o}$ derived
as above, then;
\begin{equation}
M_{o} = 5log(\overline{10^{0.2M_{o}}}) - 0.23\sigma_{1}^{2}.
\end{equation}
Thus for the case of a constant underlying space density the best estimate
of $M_{o}$ is given by;
\begin{equation}
M_{o} = 5log(\overline{10^{0.2M}}) + 1.151\sigma_{t}^{2} - 0.23 \sigma_{1}^{2}
\end{equation}

Note that the coefficient of $\sigma_{t}^{2}$
in this equation is different from that in the conversion
between $M_{o}$ and $M_{m}$, the mean value for objects of a given $m_{o}$
(=1.38). This latter factor is the well-known Malmquist (1920) bias, first
given explicitly by Eddington (1914).

In the case of the Hipparcos parallaxes of Cepheids discussed by FC, the
bias terms in equation 8 are negligibly small. The total bias correction is
0.010mag and this would change the PL zero-point from --1.43 to --1.42. The
very small bias correction is due to the method of analysis adopted by FC.
This was to reduce the intrinsic scatter about the PL relation by
the combined use of a PL and PC relation for reddening correction and
the determination of relative absolute magnitudes. If this is not done
it might be necessary to consider whether Cepheids are distributed
uniformly throughout a strip in the PL plane, at least at long periods.
In that case, and assumimg $\sigma_{m_{o}}$ is small and a constant space 
density distribution, equation 5 above would need to be replaced by;
\begin{equation}
\overline{e^{b\epsilon}} = 3sinh(2b\Delta)/2sinh(3b\Delta)
\end{equation}
where $\Delta$ is the halfwidth of the strip in magnitudes. If one
adopts $2\Delta \sim 0.7$mag at $V$ from the longer period LMC Cepheids
(Caldwell \& Coulson 1986), the correction factor would be
$\sim +0.05$mag.

The above considers the case when the objects have a mean absolute magnitude
per unit volume of $M_{o}$ with a gaussian scatter $\sigma_{M_{o}}$.
In a number of cases the relative absolute magnitudes of the objects
concerned are known through the measurement of some auxilliary quantity, $y$,
and a relation of the form;
\begin{equation}
M_{o} = Ay + B
\end{equation}
where A is a known constant. Two separate cases then arise. if the measuring
error in $y$, $\sigma_{y}$, introduces an error in $M_{o}$ which is
negligibly small compared with the intrinsic dispersion in $M_{o}$ at a 
given $y$, then the results given above apply in this case with obvious
modifications to equation 1 (see Feast 1987). If this is not the case,
then the quantities $v(x)$ etc. have to be replaced by $P(x)$ etc.,
where $P(x)$ is the distribution in $x$ of the objects actually under
discussion (i.e. taking into account that only a fraction, $f(x)$, of all
the objects may have been observed at a given $x$ (see Feast 1972). 
This procedure
essentially involves using the inverse solution of equation 10. In the
case of the Hipparcos Cepheids, the quantity $y$ is the period and this
is very accurately known, so that the earlier discussion holds in this case.

Even in the case of Mira variables, the percentage uncertainty in the periods
probably makes a much smaller contribution to the uncertainty in $M_{o}$
than the intrinsic dispersion. Thus equation 8 can be applied to the 
discussion of the Hipparcos parallaxes of these stars by Whitelock \&
Feast (2000). These workers found an infrared PL zero-point of
$+0.84 \pm 0.14$. Using their data one obtains from equation 8 a correction
of $\sim 0.02$mag, making the zero-point $+0.86 \pm 0.14$.

There is another source of bias that should be discussed. In an analysis 
it is necessary to work with $m_{o}$, the absorption free absolute 
magnitude. Evidently if absorption is present, objects of a given $m_{o}$
will have a range of uncorrected apparent magnitudes and selection
according to apparent magnitude will affect the bias correction.
In general absorption increases with distance. Thus if the fraction of
objects of a given apparent magnitude observed decreases with increasing 
apparent
magnitude, this will have the effect of reducing the fraction of
objects of a given $x$ whose true distance modulus is greater than
($m_{o} - M_{o}$). Thus absorption, if it affects the bias, is likely to do
so by reducing the number of objects in the sample whose lumniosities
are greater than the average at a given $m_{o}$. This means for instance that,
if anything, the coefficient of the second term of the r.h.s of equation 8
will need decreasing. Thus the value of $M_{o}$ produced from an
unchanged equation 8 would be too faint.

\section{Observations of Individual Objects}
Provided there is no systematic error in the measuring process, an observed
parallax is not in itself biased. This was, in fact, clearly recognized
by Lutz \& Kelker (1973) who point out that the distribution of measured 
parallaxes about the true parallax is expected to be gaussian. So provided
the objects concerned are not chosen or weighted by their measured $\pi$
or by $\sigma_{\pi}/\pi$, an analysis is not subject to bias of the
Lutz-Kelker type. This has been stressed recently by e.g. Whitelock \&
Feast (2000) and Groenewegen \& Oudemaijer (2000). 

It is of interest to consider the case when only one object of a class is
measured, as in the important Hubble Space telescope (HST) observations of
the parallaxes of RR Lyrae and $\delta$ Cephei (Benedict et al. 2002a,
Benedict et al. 2002b). These stars were presumably chosen as bright
members of their class and their HST parallaxes subsequently measured.
The absolute magnitudes derived from the parallaxes are therefore not
subject to Lutz-Kelker bias. However, since the method of determining
an absolute magnitude from a single object is effectively the method of
reduced parallaxes applied to one member alone of the class, the bias
corrections outlined above will be required, i.e. for the case of an
adopted uniform space distribution of the objects of the class, equation 8,
if an estimate of $M_{o}$ is required. If the best estimate of the absolute
magnitude of the individual object itself is wanted then, obviously,
the term in $\sigma_{t}$ in equation 8 is not required. 

In the case of $\delta$ Cephei (Benedict et al. 2002b) and adopting their
parallax, apparent magnitude and reddening, and with
$\sigma_{M_{o}} = 0.21$ (Caldwell \& Coulson 1986), $\sigma_{1} = 0.1$
and assuming $\sigma_{m_{o}}$ can be neglected one obtains an estimate of
$M_{o}$ of $-3.41\pm 0.10$ and a zero-point of the PL relation adopted
by FC of $-1.36 \pm 0.10$. Alternatively one can use the PL and PC relations
together for the reasons discussed above to obtain the reddening and
absolute magnitude. This gives $M_{o} = -3.37 \pm 0.10$ and a PL zero-point
of $-1.32 \pm 0.10$ which may be compared with the FC zero-point
(corrected as above) of $-1.42 \pm 0.12$ (where the standard error of FC
has been increase from $0.10$ for reasons given in Feast 1999). Evidently the
agreement is good, as it is with other Cepheid zero-point estimates (see
e.g. Feast 2002). Note that, although uncertainty in the reddening is a
limiting factor in these determinations of absolute magnitude, it is not
important in the use of Cepheids as distance indicators so long as
a consistent reddening scale is used for both calibrators and programme stars.
This is most easily accomplished using a standard PC relation. The
application of the Cepheid scale to derive the distance of the LMC is
made somewhat uncertain by the need for rather poorly known metallicity
corrections. However, this is not a problem in their use in the
HST ``key'' programme on extragalactic Cepheids (Freedman et al. 2001) since
the weighted mean metallicity of these galaxies is close to solar 
(Feast 2001).

In the case of RR Lyrae, using the adopted parallax, apparent magnitude 
and reddening from Benedict et al. (2002a) and assuming RR Lyrae
variables of the relevant metallicity ([Fe/H] = --1.39) uniformly fill
an instability strip of width $\sim 0.4$mag, as is the case in
globular clusters (see Sandage 1990, especially fig. 16), one obtains
as an estimate of the absolute magnitude of RR Lyrae variables
at an [Fe/H] of --1.39, $M_{o} = +0.64 \pm 0.11$. It has of course to be
realized that, especially in the case of single objects, the uncertainty
in any applied bias corrections is large.

Obviously the same general arguments apply if more than one object of a class 
is observed. Bias of the Lutz-Kelker type arises when the individual 
absolute magnitudes of objects derived from parallaxes are combined by
weighting them using the observed parallaxes. The effect of this on the 
weighting is illustrated in a practical case in Feast (1998). In some at least 
of these cases there will also be selection by apparent magnitude, so that
corrections such as that given by equation 8 will apply in addition to
the Lutz-Kelker correction.

\section{Conclusion}
The bias corrections to the absolute magnitude scales of Cepheids, RR Lyraes
and Miras discussed in this paper are all relatively small. However,
with the forthcoming great improvements in parallax measurements expected
from GAIA and other space missions, bias corrections of this type will
become increasingly significant relative to other sources of uncertainty.

\section*{Acknowledgments}
I am grateful to the referee (Dr G.F. Benedict) for his comments.

\end{document}